%% file: sample-sigconf.tex
\documentclass[sigconf,nobookmarks=true, 9pt]{acmart}
\copyrightyear{2025}
\acmYear{2025}
\setcopyright{acmlicensed}
\acmConference[ISPD '25] {Proceedings of the 2025 International Symposium on Physical Design}{March 16--19, 2025}{Austin, TX, USA.}
\acmBooktitle{Proceedings of the 2025 International Symposium on Physical Design (ISPD '25), March 16--19, 2025, Austin, TX, USA}
\acmISBN{979-8-4007-1293-7/25/03}
\acmDOI{10.1145/3698364.3705347}

\usepackage{graphics}
\usepackage{graphicx}
\usepackage{amsfonts}
\usepackage{amsmath}
\usepackage{multirow}
\usepackage{xcolor}
\usepackage{lipsum}
\usepackage{subfig}
\usepackage{algorithm}
\usepackage{algpseudocode}
\algtext*{EndWhile}
\algtext*{EndIf}
\algtext*{EndFor}
\algtext*{EndProcedure}
\usepackage[super]{nth}
\usepackage[normalem]{ulem}

\newcommand{\blue}[1]{\textcolor{blue}{#1}} 

\newtheorem{problem}{Problem}

\begin{document}

\title[\footnotesize{DRC-Coder}]{DRC-Coder: Automated DRC Checker Code Generation \\ Using LLM Autonomous Agent}


\author{Chen-Chia Chang}
\affiliation{%
  \institution{Duke University}
  \city{Durham}
  \state{NC}
  \country{USA}
}
\email{chenchia.chang@duke.edu}

\author{Chia-Tung Ho}
\affiliation{%
  \institution{NVIDIA Research}
  \city{Santa Clara}
  \state{CA}
  \country{USA}
}
\email{chiatungh@nvidia.com}

\author{Yaguang Li}
\affiliation{%
  \institution{NVIDIA}
  \city{Austin}
  \state{TX}
  \country{USA}
}
\email{yaguangl@nvidia.com}

\author{Yiran Chen}
\affiliation{%
  \institution{Duke University}
  \city{Durham}
  \state{NC}
  \country{USA}
}
\email{yiran.chen@duke.edu}

\author{Haoxing Ren}
\affiliation{%
  \institution{NVIDIA Research}
  \city{Austin}
  \state{TX}
  \country{USA}
  }
\email{haoxingr@nvidia.com}

\keywords{design rule checking, code generation, large language model}

\ccsdesc[500]{Hardware~Physical design (EDA)}
\ccsdesc[500]{Computing methodologies~Artificial intelligence}


\begin{abstract}

In the advanced technology nodes, the integrated design rule checker (DRC) is often utilized in place and route tools for fast optimization loops for power-performance-area.
Implementing integrated DRC checkers to meet the standard of commercial DRC tools demands extensive human expertise to interpret foundry specifications, analyze layouts, and debug code iteratively.
However, this labor-intensive process, requiring to be repeated by every update of technology nodes, prolongs the turnaround time of designing circuits. 

In this paper, we present DRC-Coder, a multi-agent framework with vision capabilities for automated DRC code generation. 
By incorporating vision language models and large language models (LLM), DRC-Coder can effectively process textual, visual, and layout information to perform rule interpretation and coding by two specialized LLMs.
We also design an auto-evaluation function for LLMs to enable DRC code debugging.
Experimental results show that targeting on a sub-3nm technology node for a state-of-the-art standard cell layout tool, DRC-Coder achieves perfect F1 score 1.000 in generating DRC codes for meeting the standard of a commercial DRC tool, highly outperforming standard prompting techniques (F1=0.631). 
DRC-Coder can generate code for each design rule within four minutes on average, which significantly accelerates technology advancement and reduces engineering costs.
\vspace{-2mm}
\end{abstract}

\maketitle

\input{txt/introduction_general}

\input{txt/preliminaries}

\input{txt/method}

\input{txt/experiment}

\input{txt/conclusion}

\bibliographystyle{ACM-Reference-Format}
\balance
\bibliography{sample-base}

\end{document}

%% file: txt/introduction_general.tex
\section{Introduction}

\begin{figure}[t]
\centering
\includegraphics[width=\linewidth]{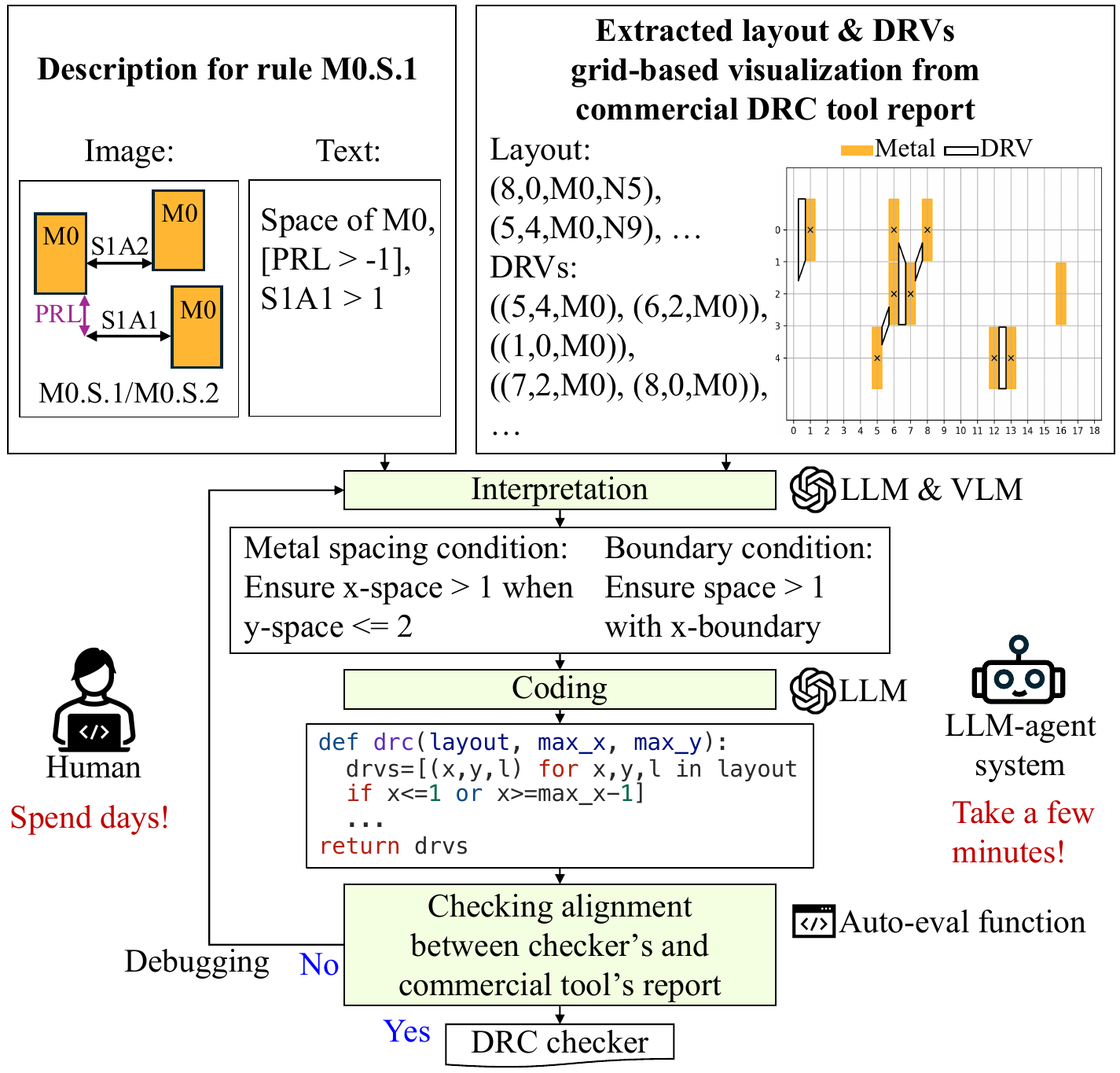}
\caption{
DRC checker development process. 
The flow begins with the interpretation of foundry-provided description and a layout with its DRVs extracted from a commercial DRC tool report. 
Then, the flow comes to coding, alignment checking, and debugging. 
The proposed LLM-agent system automates this process, significantly reducing the development time compared to manual coding.
}
\label{fig:motivation}
\end{figure}

In the era of advanced technology nodes, design rule checking (DRC) is a critical yet complex step in physical design due to the increasing number of design rules, more complex inter-layer design rules, and strict patterning rules.
Place and route (P\&R) tools often require an integrated DRC checker to ensure manufacturability and enable faster optimization loops for power-performance-area (PPA) than running commercial DRC tools every iteration. 
Implementing an integrated DRC checker typically takes experienced engineers several weeks, involving numerous iterations of debugging to extract DRC rules from foundry documents and ensure the integrated checker meets the standards of commercial DRC tools, as shown in Figure~\ref{fig:motivation}.
Furthermore, this process need to be repeated for each new technology development, which significantly cause the long turnaround time of designing circuits. 
Thus, an efficient and intelligent methodology for DRC checker code generation is essential to improve the consistency between the integrated DRC checker and the commercial DRC tool and reduce the development time for new technologies.

Nowadays, large language models (LLMs)~\cite{radford2019language, raffel2020exploring, chung2022scaling, nijkamp2023codegen2, gpt4o} have demonstrated remarkable reasoning and code generation capabilities.
In addition, vision language models (VLMs)~\cite{zhang2024vision, abdin2024phi, gpt4o} have been able to effectively perform multi-modal reasoning.
Moreover, LLMs have shown great potential in solving complex tasks through LLM autonomous agent (LLM-agent)~\cite{wang2024survey, yao2022react, langchain, wu2023autogen}.
For example, LLM-agent can stabilize the LLM coding process~\cite{yang2024swe} through auto-debugging.
In LLM-agent, LLMs can decompose the task, generate intermediate instructions, and provide feedback. 
In addition, LLMs can interact with external environment like calling utility functions to solve problems.
As a result, incorporating LLM-agent with VLMs and LLMs serves as a good candidate for DRC interpretation and code generation to enable auto-reasoning and auto-debugging.

In this work, we propose DRC-Coder, a multi-agent framework equipped with vision capability for automated DRC code generation.
This approach can absorb information from multiple modalities, including textual descriptions, visual illustrations, and layout representations, for comprehensive design rule interpretation. 
Our multi-agent framework breaks down the process into two hierarchical sub-tasks including interpretation and coding, mimicking the human DRC coding process shown in Figure~\ref{fig:motivation}.
By assigning two LLMs with different roles to handle each sub-task, we enhance the LLM reasoning ability and reduce the potential hallucination of LLMs in solving complex DRC coding with one agent. 
Additionally, by integrating our proposed domain-specific utility functions, DRC-Coder can perform auto-evaluation by executing the generated code on a layout dataset to get a performance report for auto-debugging, ensuring the effectiveness of the generated code.


To the best of our knowledge, we are the first work on the
automated DRC code generation problem in EDA. 
This work differs from a related work~\cite{zhu2023drc}, which only focuses on extracting key design rule components rather than generating completed codes without any human intervention.
To demonstrate the concept, DRC-Coder targets on the DRC checker for standard cell layout automation~\cite{li2019nctucell, van2019bonncell, lee2020sp, ren2021nvcell, ho2023nvcell} using a sub-3nm technology node.
This checker operates on a grid-based layout format, which represents layout metals as grid coordinates, e.g., (8, 0, M0, N5), (5, 4, M0, N9) in Figure~\ref{fig:motivation}. 
This is because routers typically perform on-track routing in the grid-based fashion.

The contribution can be summarized as follows:
\begin{itemize}
\item We present DRC-Coder, the first automated DRC checker code generation system to accelerate technology migration process and reduce engineering efforts. 
\item We develop a novel multi-agent framework with vision capability, which effectively interprets multi-modal information of design rules and layouts and enables automated debugging and feedback mechanisms.
\item We decompose DRC code generation into two hierarchical tasks, interpretation and coding, and allocate them to two specialized agents to improve LLM reasoning and reliability.
\item We propose three domain-specific utility functions for LLM-agent, including visual analysis for design rule and layout and automated code evaluation.
\item Evaluation using sub-3nm technology node show that our DRC-Coder successfully generates correct codes (F1=1.000) for all design rules considered in NVCell~\cite{ren2021nvcell}, while standard prompting produces unsatisfactory results (F1=0.631).
Additionally, DRC-Coder efficiently generates code per rule within four minutes on average.
\end{itemize}

DRC-Coder has potential to be extended to other DRC-related problems including DRC document explanation, test pattern generation, and design rule optimization.
In addition, we hope this work can pave the way to new research directions for automating complex engineering tasks in the semiconductor industry.

%% file: txt/preliminaries.tex
\section{Preliminaries}
In this section, we first study the related works for LLM-agent frameworks.
Then, we introduce VLMs and its potential on explaining design rule images and layouts.
Finally, we introduce the grid-based DRC checker used in the standard cell layout tool.

\subsection{LLM-Agent Framework}
LLM autonomous agents (LLM-agents)~\cite{wang2024survey, yao2022react, langchain, wu2023autogen} have emerged as powerful tools that enable LLM to make plans and execute external functions based on their reasoning processes.
LLM-agents have demonstrated their effectiveness in various domains. 
In online shopping scenarios~\cite{yao2022webshop}, LLM can base on user instruction to search, choose the product on the website, and reason when to buy the product to satisfy user requirements.
In programming tasks~\cite{chen2022codet}, LLM-agents can generate code, compile and execute it, and iteratively debug based on compiler and execution feedback. 
This ability to reason, act, and learn from feedback illustrates the enhanced problem-solving capabilities of LLM-agents.
In chip designs, LLM-agents are also applied for tasks like Verilog and layout clustering generation~\cite{ho2024large, ho2024verilogcoder}, demonstrating their potential in specialized domains and hardware-related problems.

However, existing frameworks only processes pure text representations, which are not effective for interpreting circuit layouts and design rules.
Therefore, having visual understanding capabilities in LLM-agent frameworks is essential.
In addition, we should provide a domain-specific DRC code evaluation function to give meaningful feedback on the code performance in detecting DRVs.
This could help LLM to effectively avoid false positives and negatives of DRVs produced by the generated code. 

\begin{figure}[t]
\centering
\includegraphics[width=\linewidth]{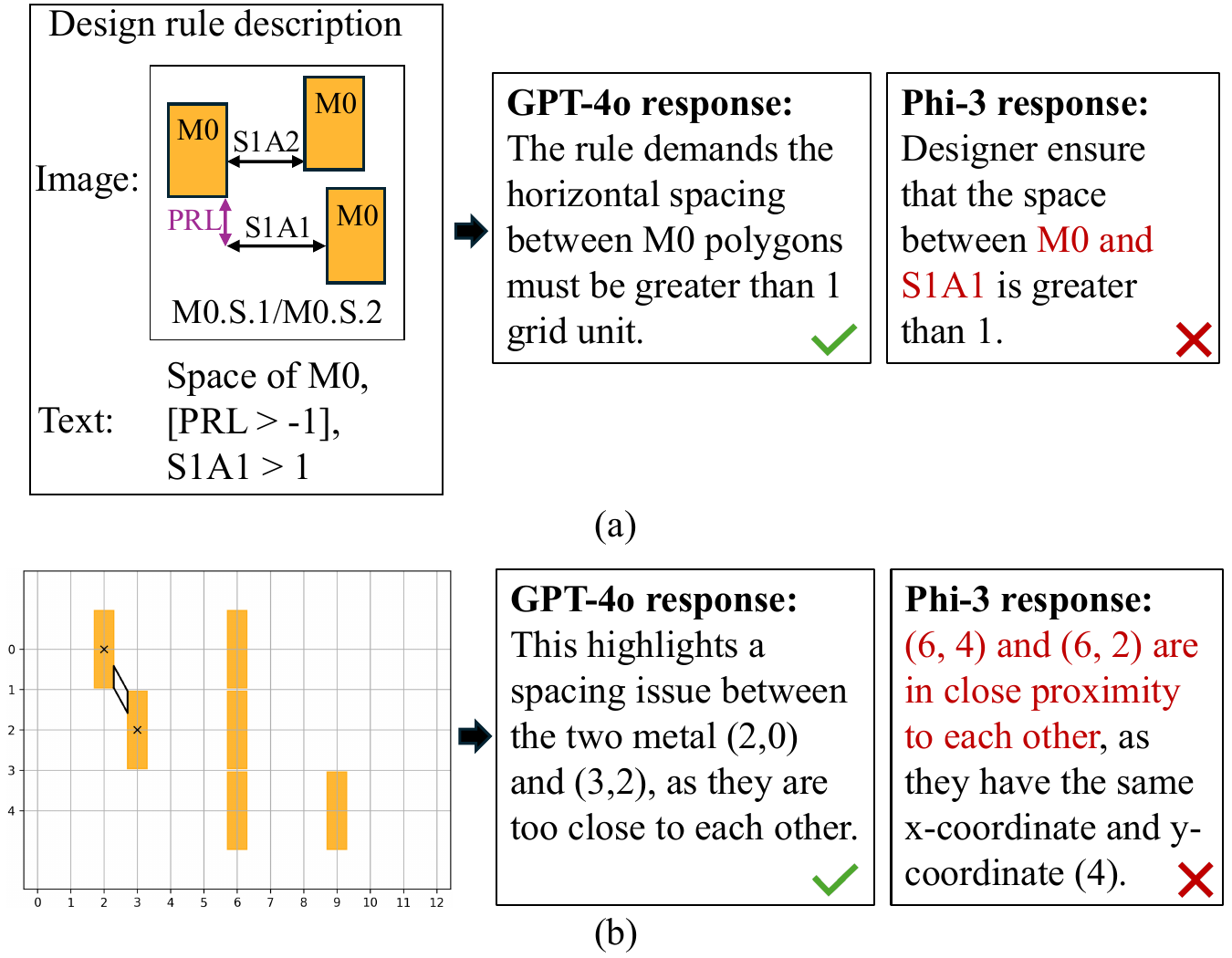}
\caption{
Comparison of the response of two VLMs, GPT-4o and Phi-3, for (a) the design rule and (b) the layout.
In layout image, yellow polygons are the metals in the M0 layer, black polygons are the DRV region marked by the commercial DRC tool, and black crosses are the corresponding DRV locations in grid-based coordinates.
}
\label{fig:exp_VLM}
\end{figure}

\subsection{DRC Interpretation Challenges \& VLMs}
Foundries specify each design rule through concise text description and visual illustration that often imply complex spatial conditions.
For example, the description of rule M0.S.1 in Figure~\ref{fig:motivation} has an image, presenting multiple spacing scenarios S1A1, S1A2 for rules M0.S.1 \& M0.S.2, and a text with abbreviated term PRL.
It needs interpretation to know that the actual condition is: the \textit{horizontal} space between metals in the M0 layer must be $> 1$ when the parallel run length (PRL) $\geq -1$, where PRL can be viewed as a \textit{vertical} space.

Circuit designers must also analyze commercial DRC tool reports on layouts to uncover implicit conditions not explicitly stated in the foundry description.
Figure~\ref{fig:motivation} reveals that the commercial tool further checks for a boundary condition: the space between the x-boundary and metals must be $> 1$.
The above example indicates the challenges of accurately analyzing foundry descriptions and commercial tool reports in DRC code generation.

VLM~\cite{zhang2024vision} can process images and text to answer the user's query based on the image.
For example, it can give image explanation or distinguish the difference between images.
Thus, VLM has the potential to help LLM-Agent to explain design rule images and layouts.
Here, we use two state-of-the-art VLMs, including Phi-3~\cite{abdin2024phi} and GPT-4o~\cite{gpt4o}, to perform image explanation on a design rule and a layout. 
The result is shown in Figure~\ref{fig:exp_VLM}. 
GPT-4o can generate meaningful responses. 
On the other hands, Phi-3 produces unsatisfied responses, where the space is not between M0 and S1A1, and the DRVs is not in (6, 4) and (6, 2).
Based on this experiment, we observe that the VLM, especially GPT-4o, can help in explaining complex design rules and layouts. 
Thus, we integrate this VLM with LLM-agents to process textual, visual, and layout information, enabling effective rule interpretation for DRC code generation.

\begin{figure}[t]
\centering
\includegraphics[width=\linewidth]{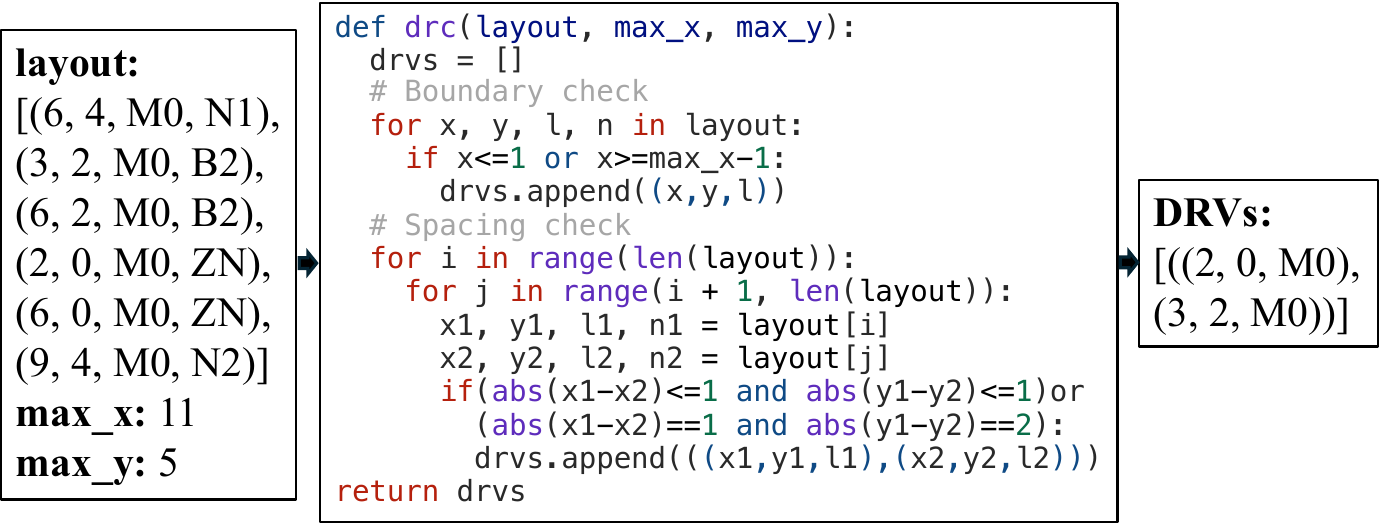}
\caption{
A grid-based DRC code for design rule M0.S.1.
}
\label{fig:grid_DRC_checker}
\end{figure}

\subsection{Grid-based DRC Checker} \label{sec:grid-based drc checker}
In our evaluation, we use NVCell~\cite{ren2021nvcell} as our target standard cell layout tool.
NVCell employs a grid-based DRC checker to rapidly obtain layout performance. 
For example, it can complete DRC for a cell with 22 devices in 0.05 seconds, while the commercial DRC tool takes 215 seconds.
In this paper, our goal is to generate a new grid-based DRC checker for a sub-3nm technology node.
Figure~\ref{fig:grid_DRC_checker} demonstrates an example of the grid-based DRC code for design rule M0.S.1 (Figure~\ref{fig:exp_VLM}(a)).
The core of the checker is the \texttt{drc} function, which takes three parameters:
\begin{itemize}
    \item \texttt{layout}: A list of tuples. Each tuple $(x, y, \text{layer}, \text{net})$ denotes the grid coordinate $(x, y)$, metal layer, and net name of a layout component. 
    Note that we use the same coordinate system for all layers, as $x$ and $y$ are horizontal and vertical coordinates, respectively.
    \item \texttt{max\_x} and \texttt{max\_y}: The maximum grid coordinates of the layout in $x$ and $y$ directions.
\end{itemize}
This \texttt{drc} function implements two  checks:
\begin{enumerate}
\item Boundary check: Identifies components at or beyond the layout edges $(x \leq 1 \text{ or } x \geq \text{max\_x}-1)$.
\item Spacing check: Detects violations between pairs of components based on their relative positions.
\end{enumerate}
The function returns a list of DRVs, where each DRV is a tuple that indicates the violating components.
In the example, the output is [((2, 0, M0), (3, 2, M0))], which indicates a spacing violation between two components (2, 0, M0) and (3, 2, M0).

%% file: txt/method.tex
\begin{figure}[t]
\centering
\includegraphics[width=\linewidth]{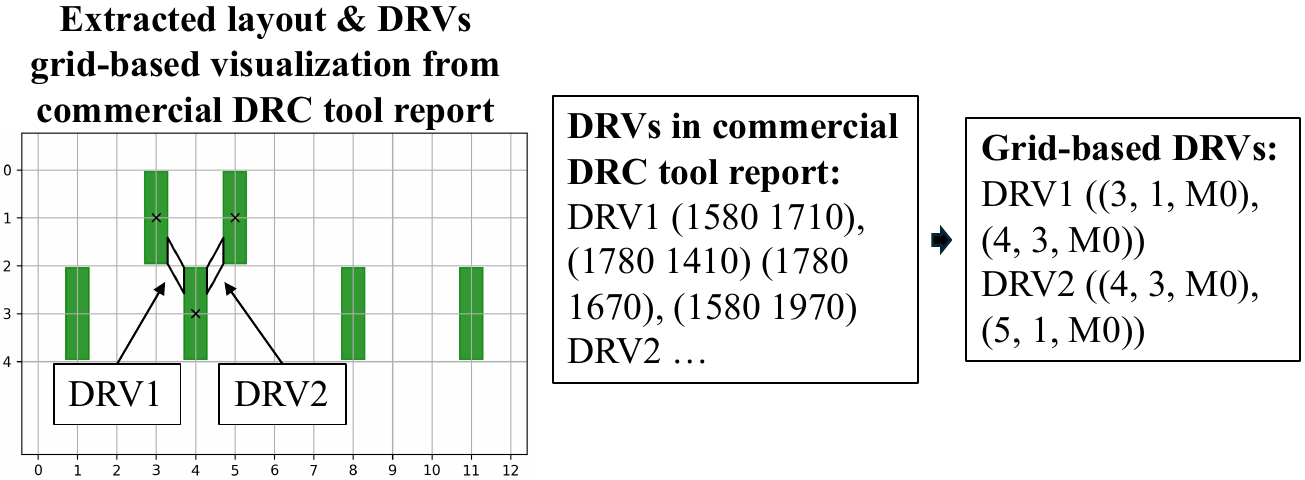}
\caption{
Conversion from DRVs in the commercial DRC tool report to grid-based DRVs.
DRV locations in the commercial tool report are marked by black polygons, each defined by four points with x and y coordinates.
Our grid-based approach identifies the layout components intersecting these polygons and represents DRVs using the grid coordinates of these components. 
}
\label{fig:conversion}
\end{figure}

\begin{figure*}[t]
\centering
\includegraphics[width=\linewidth]{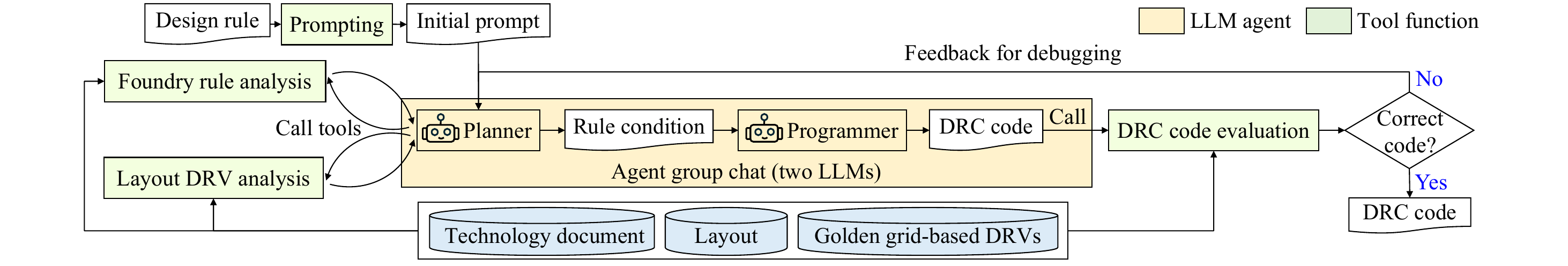}
\caption{Overview of DRC-Coder. 
Planner first interprets the input design rule by executing analysis tool functions.
Programmer receives the rule condition to generate code. 
Finally, DRC code evaluation is executed to provide code performance feedback. 
Planner receives the feedback to perform re-reasoning, and Programmer performs debugging until generating the correct code.
}
\label{fig:flow}
\end{figure*}

\begin{figure}[t]
\centering
\includegraphics[width=\linewidth]{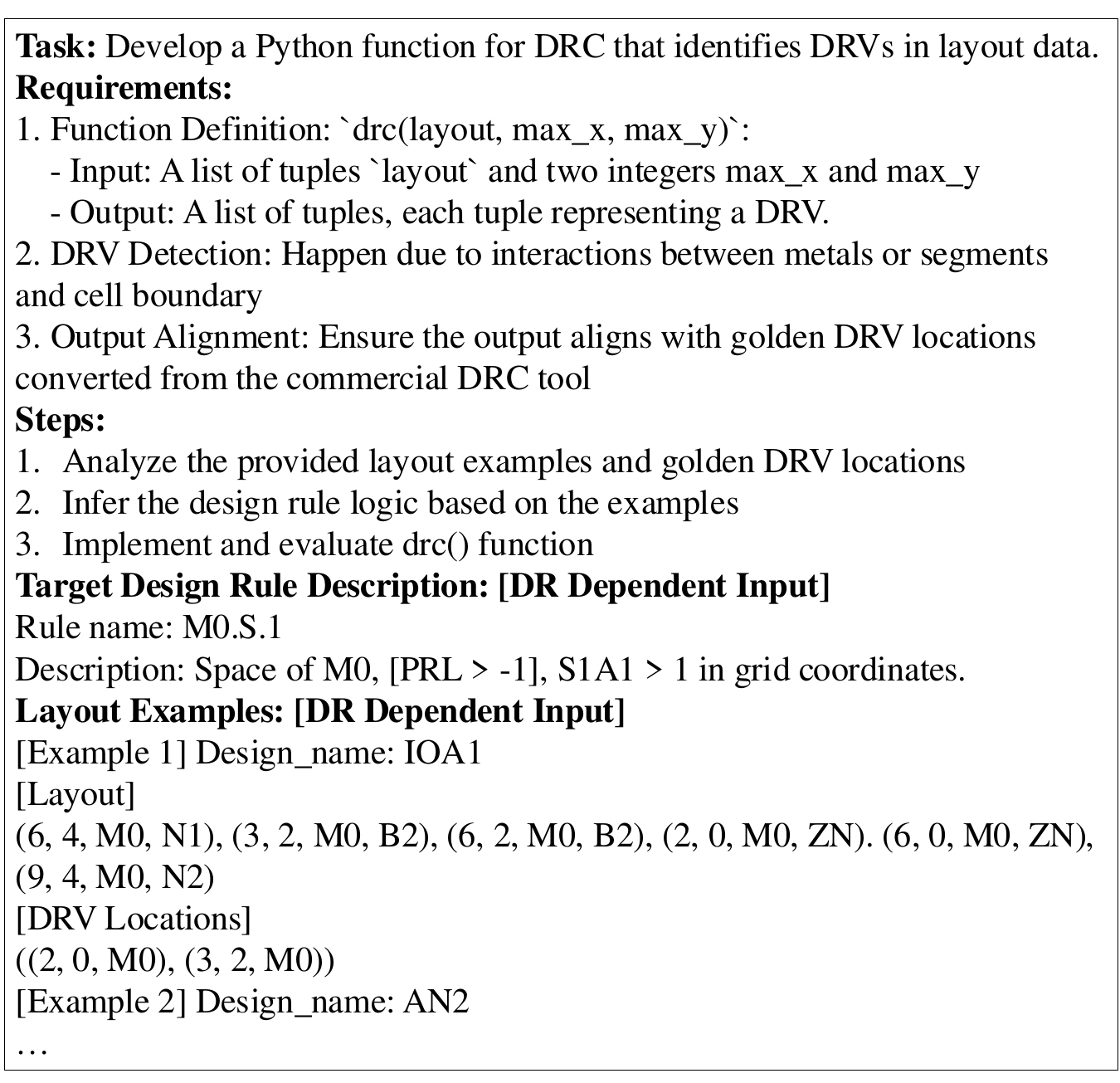}
\caption{
Initial prompt to DRC-Coder. The part of design rule (DR) dependent inputs is dynamically changed according to different target design rules.
}
\label{fig:initial_prompt}
\end{figure}

\section{Data Preparation} \label{sec:data_prepare}
To evaluate the generated DRC code, we create a dataset composing of standard cell layouts and their DRC reports. The dataset preparation process has two steps: layout generation and DRC report preprocessing.

\subsection{Layout Generation}

We produce 207 different standard cell layouts using NVCell by mutating the routing behaviors without DRC fixing. 
This approach ensures a wide range of DRV scenarios for evaluation. 
These layouts are represented in a grid format used in the grid-based DRC checker stated in Section~\ref{sec:grid-based drc checker}.

\subsection{DRC Report Preprocessing}
The preprocessing stage converts physical coordinate-based DRC reports from the commercial tool into a grid-based representation that aligns with the output format of our grid-based DRC checker.
The preprocessing involves the following steps:
\begin{enumerate}
    \item Produce the DRC reports of layouts by running the commercial DRC tool. These reports use polygons to mark DRV locations in physical coordinates.
    \item Identify the layout components that intersect with the DRV polygons reported by the commercial DRC tool. 
\end{enumerate}


Finally, we view the grid-based coordinates of these layout components as the ground-truth of DRVs in our evaluation process.
This grid-based DRV report can precisely capture layout components involved in each DRV.
Figure~\ref{fig:conversion} illustrates this conversion process, demonstrating how polygon-based DRVs from the commercial tool reports are transformed into our grid-based representations.
For the DRV interpretation in DRC-Coder, we also construct the grid-based visualization of each layout along with its DRVs, as shown in the examples in Figure~\ref{fig:motivation}, \ref{fig:exp_VLM}(b), and \ref{fig:conversion}.

\section{DRC-Coder}

Our approach, DRC-Coder, generates DRC code on a rule-by-rule basis.
To make DRC-Coder more applicable when facing new technology node, our code generation is under a zero-shot setting, i.e., no example codes are provided during generation.
The overall flow of DRC-Coder is illustrated in Figure~\ref{fig:flow}.
The core of this system consists of two LLM agents operating in a group chat manner: 
(1) Planner: Responsible for interpreting design rule conditions in grid domain. 
(2) Programmer: Translate the design rule condition into the executable code. 
These agents are powered by the general LLMs (GPT-4o~\cite{gpt4o}) but are assigned with different roles for the DRC coding process.
This multi-agent approach decomposes the DRC process into planning and coding phases, allowing each agent to focus on its specialized task to enhance overall system performance.

To process image inputs and evaluate DRC code, DRC-Coder sets three specialized tool functions for agents to use: (1) Foundry Rule Analysis, (2) Layout DRV Analysis, and (3) DRC Code Evaluation.
These tools provides the agents the specialized image analysis and evaluation capabilities throughout the code generation process.

The workflow begins with an input design rule, which goes through Prompting stage to produce an initial prompt to Planner.
Then, Planner and Programmer work together to generate the DRC checker code with the help of tool functions.
Finally, the code generation undergoes an iterative auto-debugging process until the DRC reports are aligned with ground truth DRVs of the commercial DRC tool. 
The example of the workflow is shown in Figure~\ref{fig:demo}.
In the following, we detail all components in DRC-Coder.

\subsection{Prompting}
Given the input design rule, this stage constructs a structured initial prompt, as illustrated in Figure~\ref{fig:initial_prompt}, to the Planner.
The components of this prompt is split into the fix part and the design rule (DR) dependent part. 
The fix part contains: (1) A task definition for developing a Python function to identify DRVs in layout data. (2) The requirements that formally states the input and output format of the function. (3) A step-by-step guide that decomposes coding problem into subtasks for the Planner and Programmer.

The DR dependent part has: (1) The target design rule description from foundry document. (2) Layout examples with metal information and corresponding DRV locations to provide concrete cases for analysis.
Note that we randomly select two layout examples that has the target DRVs from our dataset to construct the prompt.
Additionally, the DR dependent part is dynamically adjusted based on the target design rule.



\begin{figure}[t]
\centering
\includegraphics[width=\linewidth]{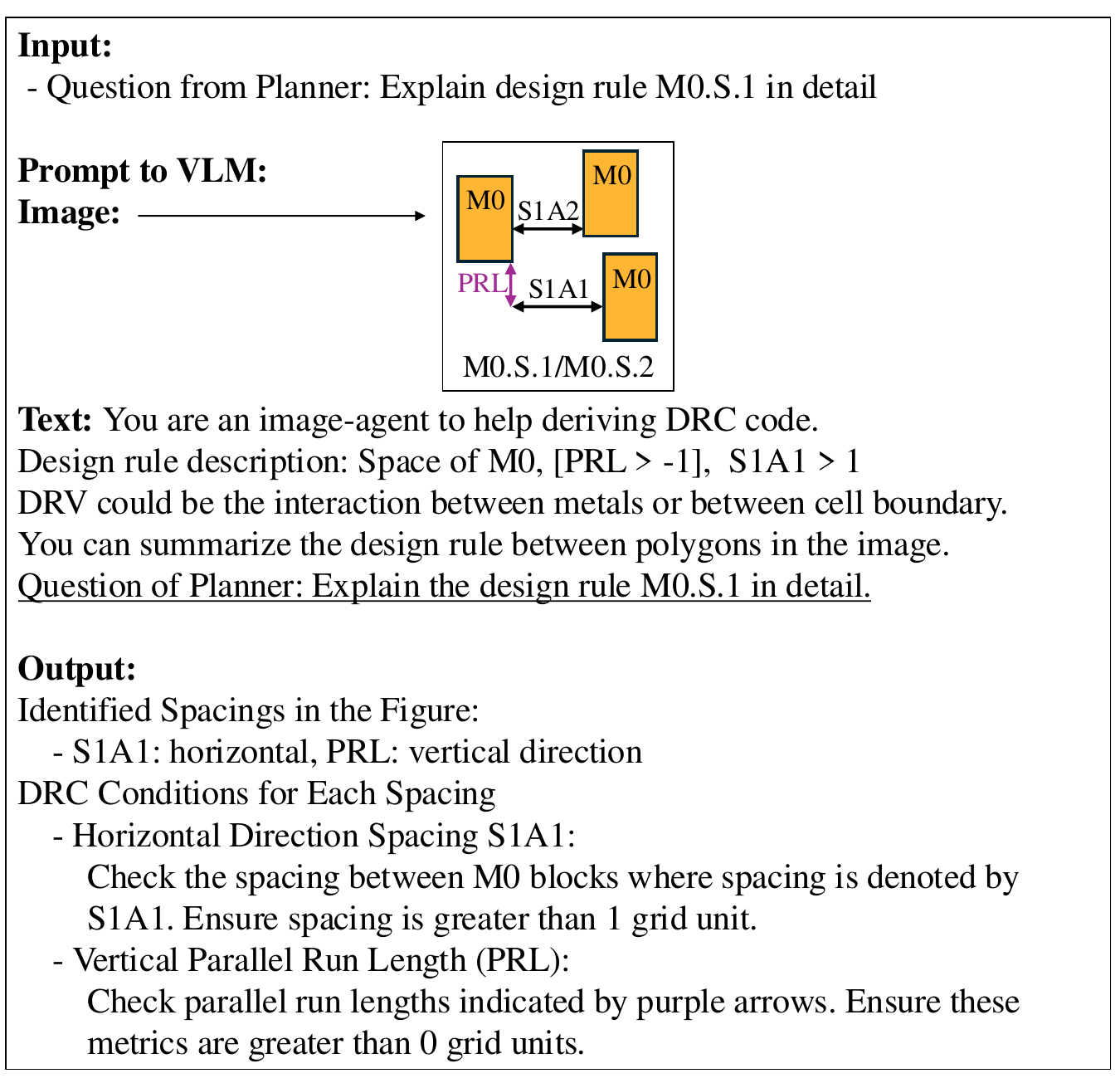}
\caption{
The usage and example response of Foundry Rule Analysis.
This function uses a VLM to interpret design rule descriptions, including text and image inputs, to produce a detailed analysis of the design rule.
}
\label{fig:foundry_explainer}
\end{figure}

\begin{figure}[t]
\centering
\includegraphics[width=\linewidth]{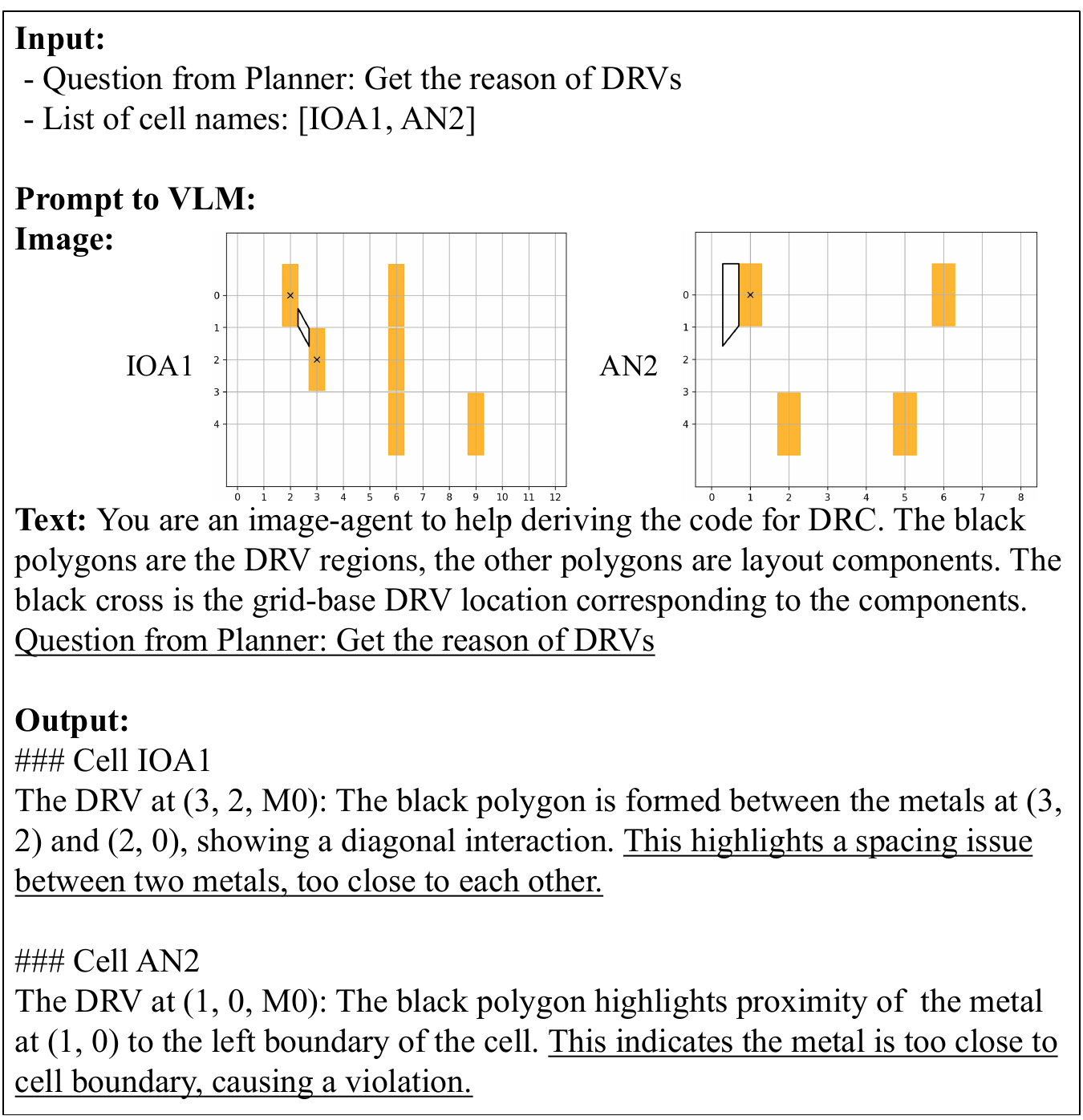}
\caption{
The usage and example response of Layout DRV Analysis. 
This function utilizes a VLM to interpret layout images of input cells, identify their DRVs, and provide detailed grid-based explanations on DRVs.
}
\label{fig:layout_explainer}
\end{figure}

\begin{figure}[t]
\centering
\includegraphics[width=\linewidth]{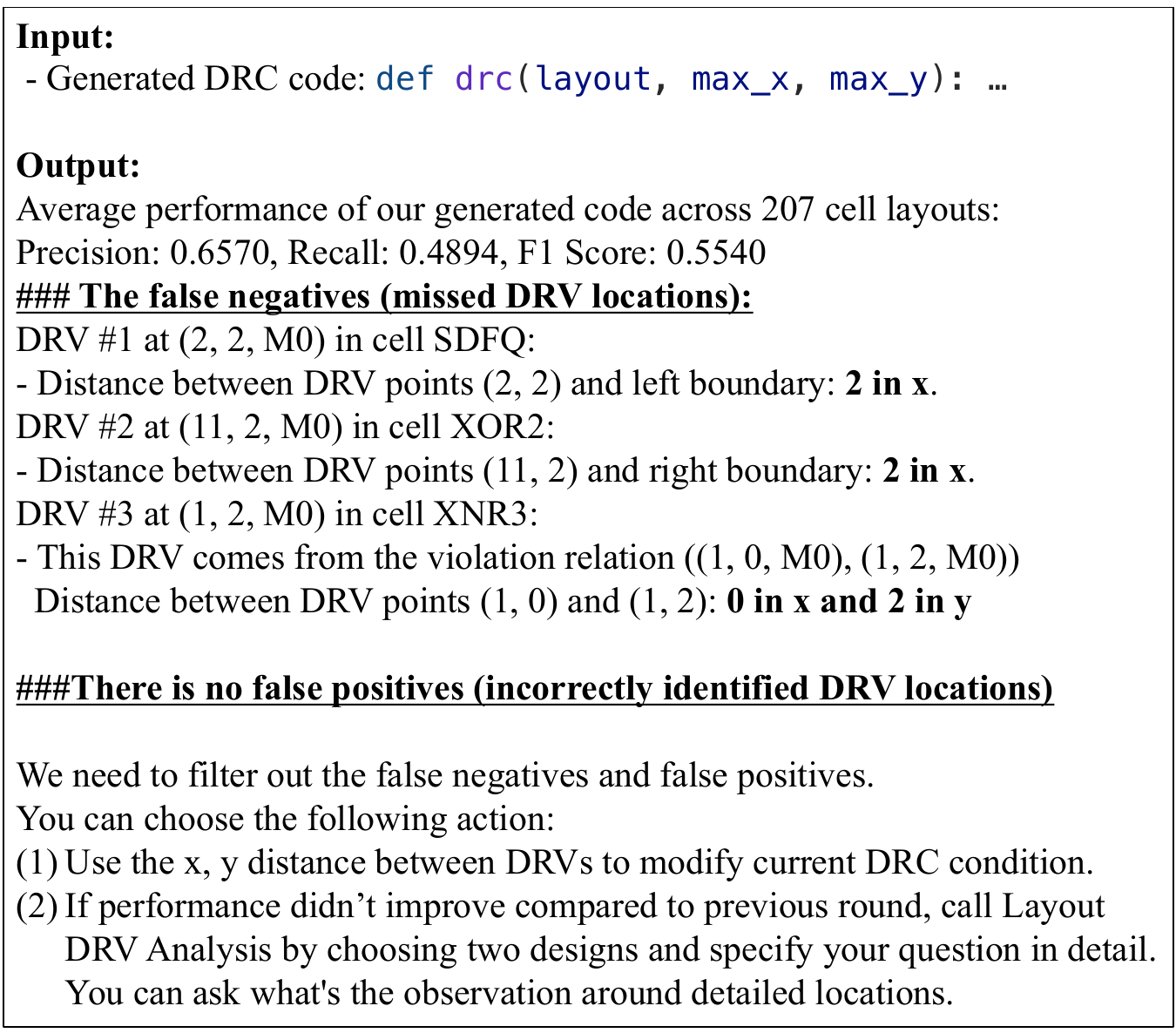}
\caption{
The usage and example response of DRC Code Evaluation.
By executing the input generated code on the layout dataset, this function compares the code outputs with golden grid-based DRVs converted from commercial tool reports. Then, it provides a performance report for our two LLM agents.
}
\label{fig:code_evaluation}
\end{figure}

\subsection{Planner}
Planner is an LLM agent focusing on interpreting foundry-provided design rule descriptions and layouts to generate corresponding design rule conditions in the grid domain. 
These foundry-provided descriptions are often concise and multi-modal, combining text and images, which makes them challenging to use directly for coding purposes. 
To help Planner interpret the image information, we design two utility functions for the Planner to employ: Foundry Rule Analysis and Layout DRV Analysis.
In each round of interpretation, Planner can control whether to call each of functions to get more information.
If calling functions, Planner receives the response of tool functions to automatically transform all information into grid-based design rule conditions.
In the example shown in Figure~\ref{fig:demo}, design rule conditions generated by Planner contain the analysis of DRVs and the plans to write the code, including boundary and spacing checks.
\\

\noindent\textbf{Foundry Rule Analysis.} 
This function processes 
a specific question from Planner regarding DRVs.
Then, a VLM is called to interpret design rule descriptions (combining text and images) in the foundry document and provides a answer to the input question.
As shown in Figure~\ref{fig:foundry_explainer}, the function analyzes the provided image, identifying target spacing directions and generating a detailed response for DRC conditions for each spacing requirement.
This automated interpretation helps Planner to understand complex design rules presented in multi-modal formats, facilitating the translation of foundry specifications into precise and grid-based conditions that can be used for DRC code generation.
\\

\noindent\textbf{Layout DRV Analysis.} 
This function takes two inputs: a question from Planner regarding design rules and layout, and a list of cell names indicating the layouts to be examined.
It then utilizes a VLM (GPT-4o) to interpret the specified layout images.
The VLM identifies key elements such as metal regions and DRV locations within the provided grid coordinates. 
As demonstrated in Figure~\ref{fig:layout_explainer}, the function generates a comprehensive response that addresses Planner's query, detailing the reasons for detected DRVs, including specific coordinates and descriptions of issues like spacing problems or boundary violations.
This automated and context analysis enhances Planner's ability to generate more accurate grid-based design rule conditions.


\subsection{Programmer}
\label{sec:Programmer}
Programmer is an LLM agent responsible for translating the grid-based design rule conditions, produced by the Planner, into executable DRC code. 
To understand the generated code performance, we design a tool function, DRC Code Evaluation.
The example in Figure~\ref{fig:demo} shows the generated DRC code.
\\

\noindent\textbf{DRC Code Evaluation.} 
This function inputs the generated code and outputs the performance report of the code. 
In detail, this function executes our generated code on cell layouts in the dataset and directly compares the code output with the golden DRC reports of the commercial tool. 
The generated DRC code must correctly classify layout grids as either DRC-compliant or DRC-violating based on each design rule. 
The dataset of standard cell layouts used for evaluation is inherently imbalanced, with DRC-violating grids being significantly less than compliant ones.
Thus, to evaluate performance, we measure Precision, Recall, and F1 score between the DRVs detected by the commercial tool and our generated code.
These metrics are particularly suitable for imbalanced datasets, focusing on the correct identification of the minority class (DRC-violating).
Higher values in these metrics indicate better DRC code.
Note that we serve F1 score as our primary metric since it offers a comprehensive view of effectiveness by balancing both precision and recall.
In this way, we can examine whether our code correctly replicates the results of the commercial tool. 
Finally, this function produces a performance report for Planner to do reasoning and for Programmer to conduct debugging.

Figure~\ref{fig:code_evaluation} shows the example of this function.
In the performance report, we have the average performance and summarize the false negatives and false positives.
For each false negatives (positives) DRV location, we first classify it into boundary violation or spacing violation. 
For each DRV of spacing violation (boundary violation), we compute the x and y distances between two points (the point to the boundary). 
Finally, we report the DRVs with unique x, y distances because they may come from different design rule conditions.
We cannot report all the DRVs due to context length limitation of LLMs and the potential hallucination problem of LLMs triggered by long contexts.

At the end of the report, we provide the goal and the available actions that agents can take.
Note that Planner has freedom to call Layout DRV Analysis in the next round of generation to know more about DRVs.
For example, Planner could ask about the DRV scenario around specific grid locations according to the report.

%% file: txt/experiment.tex
\begin{table*}[!htp]\centering
\caption{
Performance evaluation of DRC code generation using standard prompting and our DRC-Coder with GPT-4o~\cite{gpt4o} and Llama3~\cite{dubey2024llama} across seven design rules. The table presents Precision (P), Recall (R), and F1 score (F) for each method. For our DRC-Coder using GPT-4o, the number of debugging iterations and runtime in seconds are also included.
}
\label{tab:main_results}
\renewcommand{\arraystretch}{1.1}
\begin{tabular}{|c |c |c |c |c |c |c |c |c |c |c |c |c | c | c | c | c | c | c | c |}
\hline \hline
LLM &\multicolumn{6}{c|}{Llama3} &\multicolumn{8}{c|}{GPT-4o} \\ \hline
\multirow{2}{*}{Rules} &\multicolumn{3}{c|}{Standard prompting} &\multicolumn{3}{c|}{DRC-Coder} &\multicolumn{3}{c|}{Standard prompting} &\multicolumn{5}{c|}{DRC-Coder} \\\cline{2-15}
&P &R &F &P &R &F &P &R &F &P &R &F & \#iteration & Runtime (sec) \\ \hline
M0.S.1 &0.841 &0.710 &0.745 &0.795 &1.000 &0.870 &0.789 &0.527 &0.620 &\textbf{1.000} &\textbf{1.000} &\textbf{1.000} &3 &325 \\ \hline
M0.S.2 &0.657 &0.489 &0.554 &0.696 &0.817 &0.722 &0.657 &0.489 &0.554 &\textbf{1.000} &\textbf{1.000} &\textbf{1.000} &2 &121 \\ \hline
VIA0.S.1 &0.646 &0.403 &0.483 &1.000 &1.000 &1.000 &0.659 &1.000 &0.769 &\textbf{1.000} &\textbf{1.000} &\textbf{1.000} &2 &133 \\ \hline
M1.S.1 &0.540 &1.000 &0.680 &1.000 &1.000 &1.000 &0.645 &0.402 &0.482 &\textbf{1.000} &\textbf{1.000} &\textbf{1.000} &3 &354 \\ \hline
M1.S.2 &0.000 &0.000 &0.000 &0.151 &0.583 &0.234 &0.582 &0.450 &0.490 &\textbf{1.000} &\textbf{1.000} &\textbf{1.000} &2 &152 \\\hline
VIA1.S.1 &0.075 &0.725 &0.133 &0.149 &1.000 &0.255 &1.000 &1.000 &1.000 &\textbf{1.000} &\textbf{1.000} &\textbf{1.000} &1 &45 \\\hline
M2.S.1 &0.220 &1.000 &0.356 &1.000 &1.000 &1.000 &0.500 &0.500 &0.500 &\textbf{1.000} &\textbf{1.000} &\textbf{1.000} &3 &343 \\\hline \hline
Average &0.425 &0.618 &0.421 &0.684 &0.914 &0.726 &0.690 &0.624 &0.631 & \textbf{1.000} & \textbf{1.000} & \textbf{1.000} &2.3 & 210 \\\hline
\hline
\end{tabular}
\end{table*}


\section{Experimental Results} \label{sec:exp}

In this section, we first detail the experiment setup. Then, we present the evaluation results and an abalation study of DRC-Coder. 
Finally, we introduce a detailed workflow of DRC-Coder.

\begin{figure}[t]
\centering
\includegraphics[width=1\linewidth]{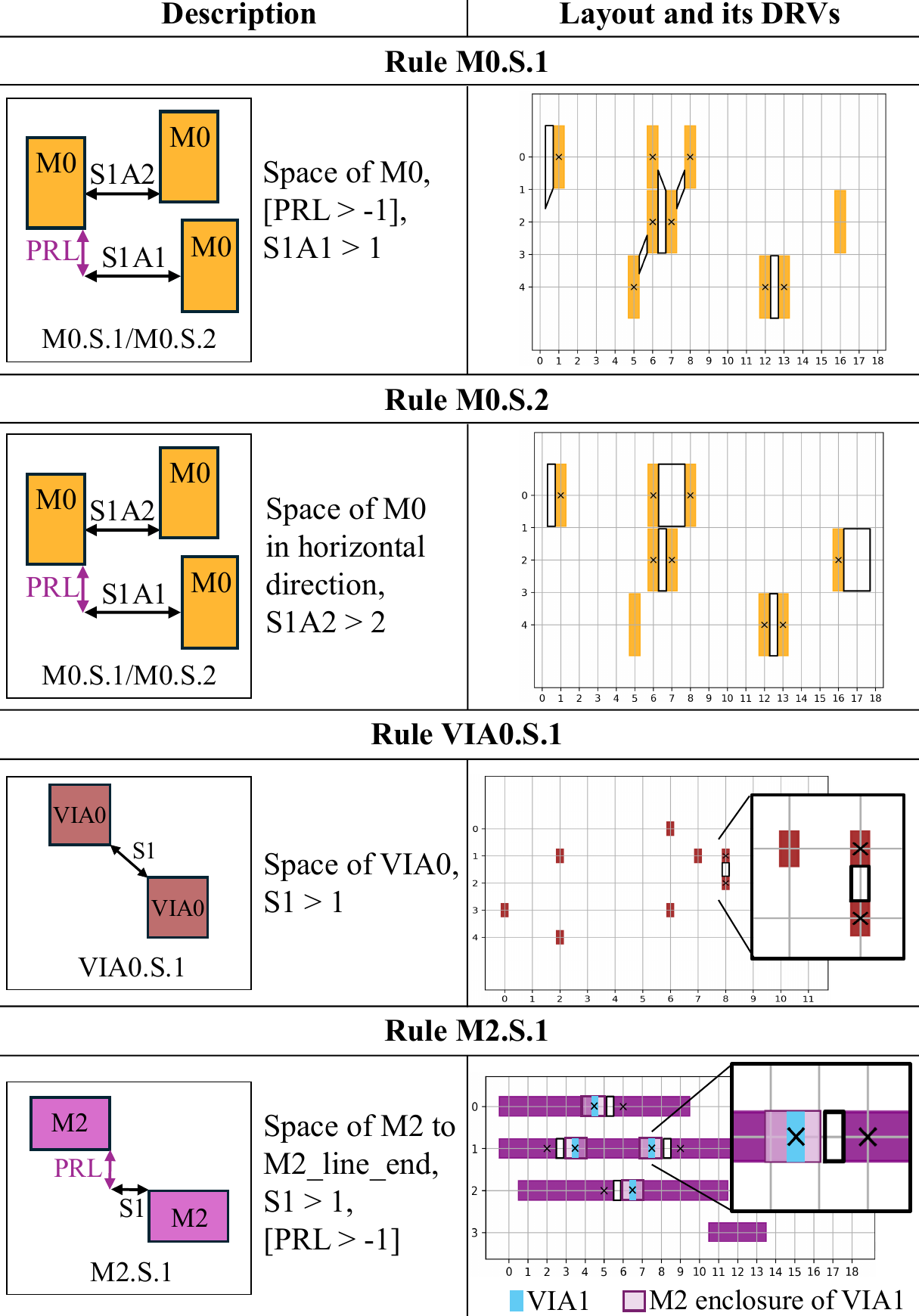}
\caption{Illustration of selected design rules for DRC-Coder evaluation. 
(1) Rule M0.S.1: Spacing between M0 components, considering both vertical and horizontal direction.
(2) Rule M0.S.2: Horizontal spacing between M0 components. 
(3) Rule VIA0.S.1: Spacing between VIA0 metals. (4) Rule M2.S.1: Interaction between VIA1 and M2 components, considering M2 enclosure of VIA1.
}
\vspace{-3mm}
\label{fig:design_rule_intro}
\end{figure}

\subsection{Experiment Setup} \label{sec:exp_setup}
\noindent\textbf{Development platform.}  DRC-Coder is developed under Python language based on the multi-agent system development toolkit AutoGen~\cite{wu2023autogen}.
Planner and Programmer agents, along with the VLMs embedded in two tool functions, are powered by GPT-4o \cite{gpt4o} using the OpenAI API version 2024-05-13. 
This means that our DRC-Coder is training-free because we do not perform any finetuning on the LLMs.
\\

\noindent\textbf{Evaluation method.}
We evaluate DRC-Coder's capability to develop DRC checker codes for a sub-3nm technology node, specifically for NVCell \cite{ren2021nvcell, ho2023nvcell}.
Our evaluation dataset, generated as described in Section \ref{sec:data_prepare}, comprises 207 standard cell layouts and covers 7 distinct design rules, as illustrated in Table~\ref{tab:main_results}.
These rules contain all primary DRV types that could occur and are considered in NVCell's grid-based routing engine.
In addition, they cover metal and via layers from M0 to M2, which include all available routing layers in standard cell layout tools \cite{li2019nctucell, van2019bonncell, lee2020sp}.

Figure~\ref{fig:design_rule_intro} illustrates four selected design rules. Rule M0.S.1 demonstrates spacing for M0 components, while rule M0.S.2 shows horizontal spacing with distinct spacing parameters.
VIA0.S.1 shows spacing constraints between VIA0 metals. 
Rule M2.S.1 represents more complex multi-layer interactions, addressing spacing between VIA1 and M2 components. 
This rule introduces the concept of M2 enclosure, which is the M2 metal that extends beyond the VIA1 boundaries, as shown by the light purple areas surrounding blue VIA squares in Figure~\ref{fig:design_rule_intro}. 
M2 enclosure is essential for ensuring reliable connections and manufacturbility in advanced chip designs.

While not illustrated, M1.S.1 and M1.S.2 are similar to M0.S.1 and M0.S.2 respectively, but with different spacing requirements in vertical directions. 
VIA1.S.1 is analogous to VIA0.S.1 but with distinct spacing parameters. 
This diverse rule set enables a comprehensive evaluation of DRC-Coder's capability to handle both single-layer spacing rules and complex multi-layer interactions.


To quantify DRC-Coder's performance, we compare the generated code output with the golden DRC reports of the commercial tool and employ three metrics: Precision, Recall, and F1 score as detailed in Section~\ref{sec:Programmer}, where F1 score is our primary metric.
\\

\noindent\textbf{Baselines.}
We are the first work focusing on DRC coding generation problem and use LLM-Agent-based method to solve it.
Thus, the main baseline is set to the standard prompting, which using prompt in Figure~\ref{fig:initial_prompt} to directly generate the code without tool function feedback.
DRC-Coder is a multi-agent framework with multi-modal vision capability.
We set two variants of DRC-Coder as other baselines for abalation study: 
(1) single-agent with vision capability: Only use Programmer to directly generate code and (2) multi-agent without vision capability: Keep Planner and Programmer but without Foundry Rule and Layout DRV Analysis.

\begin{table}[t]\centering
\caption{Performance evaluation of DRC code generation using two DRC-Coder variants:  multi-agent without vision capability and single-agent with vision 
capability using GPT-4o. P: Precision, R: Recall, F: F1 score} 
\label{tab:abalation_drc}
\renewcommand{\arraystretch}{1.1}
\begin{tabular}{|c |c |c |c |c |c |c |}\hline\hline
LLM & \multicolumn{6}{c|}{GPT-4o} \\ \hline
\multirow{3}{*}{Design rules} &\multicolumn{3}{c|}{Multi-agent} &\multicolumn{3}{c|}{Single-agent} \\
&\multicolumn{3}{c|}{w/o vision capability} &\multicolumn{3}{c|}{w/ vision capability}\\\cline{2-7}
&P &R &F &P &R &F \\\hline
M0.S.1 &0.951 &0.657 &0.747 &0.944 &0.965 &0.946 \\\hline
M0.S.2 &0.929 &1.000 &0.956 &0.530 &1.000 &0.661 \\\hline
VIA0.S.1 &1.000 &1.000 &1.000 &0.660 &1.000 &0.769 \\\hline
M1.S.1 &0.848 &0.880 &0.844 &1.000 &1.000 &1.000 \\\hline
M1.S.2 &1.000 &1.000 &1.000 &1.000 &1.000 &1.000 \\\hline
VIA1.S.1 &1.000 &1.000 &1.000 &1.000 &1.000 &1.000 \\\hline
M2.S.1 &1.000 &1.000 &1.000 &1.000 &1.000 &1.000 \\\hline
\hline
Average & 0.961 & 0.934 & 0.935 & 0.876 & 0.995 & 0.911 \\\hline
\hline
\end{tabular}
\end{table}

\begin{figure*}[t]
\centering
\includegraphics[width=1\linewidth]{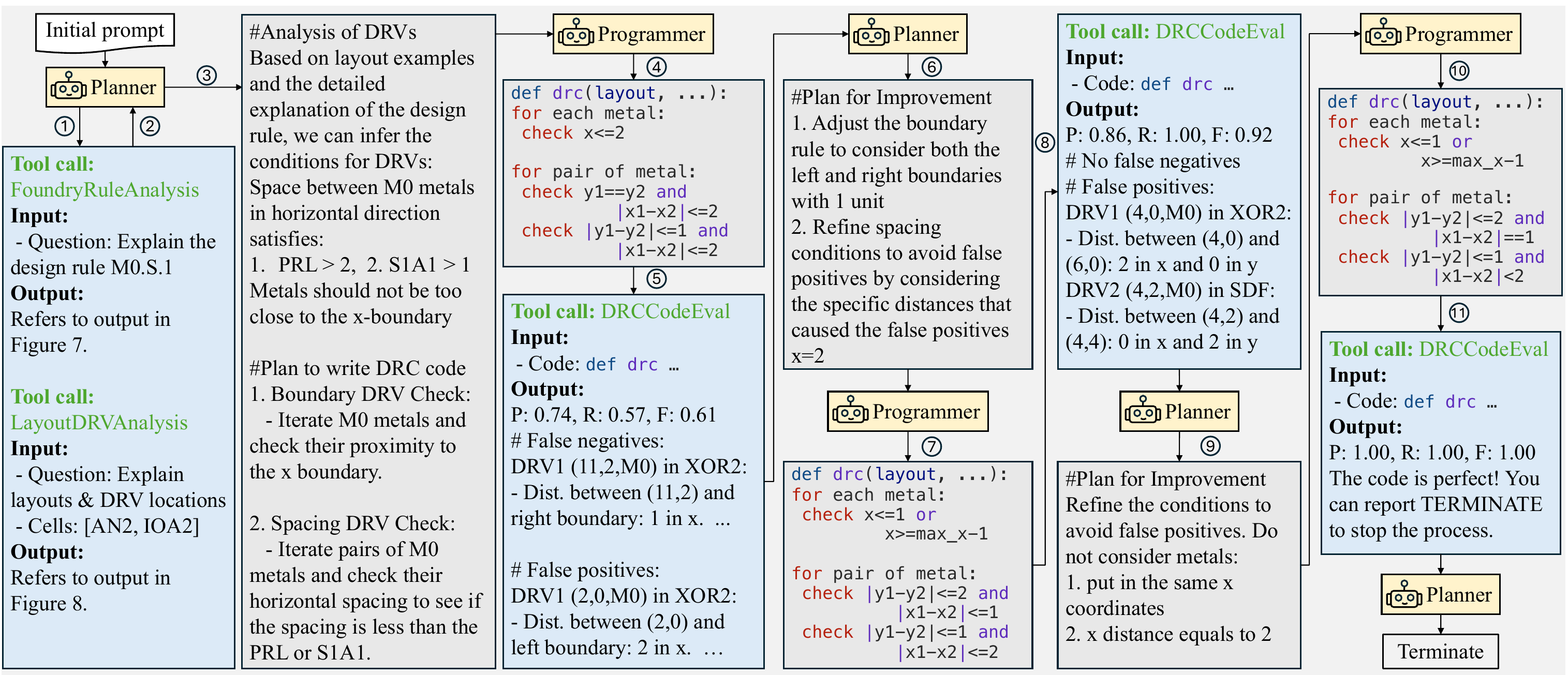}
\caption{
Workflow of DRC-Coder generating and refining DRC code for design rule M0.S.1. 
The process illustrates the iterative collaboration between Planner and Programmer, utilizing various tool calls (FoundryRuleAnalysis, LayoutDRVAnalysis, DRCCodeEval) to progressively improve code performance and eliminate false positives and negatives. Note that pseudo code is used in the image to represent the generated code to reduce the context length.
}
\vspace{-2mm}
\label{fig:demo}
\end{figure*}

\subsection{Results of DRC-Coder}
The evaluation results are shown in Table~\ref{tab:main_results}.
Our DRC-Coder using GPT-4o~\cite{gpt4o}, employing a multi-agent architecture with vision capability, achieves perfect scores (1.000) in Precision, Recall, and F1 score for all seven design rules evaluated. 
This consistent performance across different rule types highlights the robustness of our approach in interpreting and translating complex design rules into accurate DRC code.

In contrast, the standard prompting method shows unsatisfied performance across different design rules with an average F1 of 0.631. 
While it performs adequately for some rules, e.g., VIA1, it struggles with others, particularly in terms of Recall and F1 scores. 
This inconsistency shows the limitations of conventional prompting when dealing with the intricate DRC.

In summary, DRC-Coder achieves 37\% higher F1 score. In addition, DRC-Coder can complete the coding within an average 2.3 iterations of debugging, taking 210 seconds of runtime.
Thus, it can greatly accelerate the DRC coding process, where a designer easily takes weeks to write a correct DRC code.

We further demonstrate the results using an open-sourced LLM Llama3~\cite{dubey2024llama}. 
Our framework also can achieve improvement with 42.2\% compared to the standard prompting.
However, it cannot perform as effective as GPT-4o, indicating GPT-4o has a more powerful agent capability in this domain-specific DRC coding problem.

\subsection{Abalation Study}

The abalation study results are shown in Table~\ref{tab:abalation_drc}.
This experiment can evaluate the performance contribution of the visual capability and the multi-agent setting in DRC-Coder.
Two variants of DRC-Coder demonstrate improved performance over standard prompting (first column in Table~\ref{tab:main_results}) with  32.5\% and 30.7\% higher F1 score, respectively.
However, they fall short of the full DRC-Coder (second column in Table~\ref{tab:main_results}) in some design rules.
These results indicate the importance of visual capability in interpreting certain design rules and the advantage of our multi-agent setting in task decomposition of the DRC coding task.

\subsection{Case Study of DRC-Coder Workflow}

This section presents a detailed case study of DRC-Coder's workflow for generating and refining DRC code for design rule M0.S.1. 
Figure~\ref{fig:demo} provides a step-by-step visualization of this process.

The workflow begins with an initial prompt, which triggers the Planner agent to analyze the design rule using  FoundryRuleAnalysis and LayoutDRVAnalysis tool functions (steps 1-2). These analyses provide insights into the rule specifications and potential DRV conditions.
Based on this information,  Planner summarizes the DRV analysis and generate a plan for writing the DRC code (step 3), including the boundary and spacing DRV checking.
Then, Programmer implements the DRC code (step 4) and call the DRCCodeEval tool (step 5) to get the code performance and reveal areas for improvement.

In the next iteration of code generation,  Planner develops a plan for refinement (step 6), indicating how to modify the boundary rules and spacing conditions.
This guides  Programmer to make code adjustments (step 7).
This iterative process continues, with each cycle improving the code performance and decrease the false negatives and positives (steps 8-10).
Note that in step 8, there is no false negatives for the code.
Finally, when the DRCCodeEval indicates the code is correct,  Planner send the TERMINATE signal to end the code generation process.

This demonstration shows that Planner can generate effective plans for modifying design rule conditions. 
Also, Programmer can follow the plan and combine its last generated code to generate an improved one.

%% file: txt/conclusion.tex
\section{Conclusion} \label{sec:conclusion}

In this work, we introduce DRC-Coder, the first automated DRC code generation framework leveraging a multi-agent system with vision capabilities. 
Our approach decomposes the DRC coding process into interpretation and programming tasks, utilizing two LLMs and integrating VLMs to effectively process multi-modal information including textual descriptions, visual illustrations, and layout representations. 
In addition, we develop three specialized tool functions for LLMs: foundry rule analysis, layout DRV analysis, and DRC code evaluation. These functions enable automated reasoning and debugging, significantly robustify the code generation process.

Our evaluation demonstrates that DRC-Coder significantly outperforms standard prompting techniques, achieving perfect F1 scores of 1.000 across all design rules considered in a standard cell layout tool for a sub-3nm technology node.
This indicates that the generated DRC checker successfully replicates the report of the commercial tool, providing signoff DRC to the layout tool.
Moreover, DRC-Coder drastically reduces the coding time from days of manual effort an average of four minutes per design rule, highly accelerating technology migration and reducing engineering costs.
Note that DRC-Coder can be generalized to generate 
codes using other programming language, e.g., C++, for more efficient DRC.

Looking ahead, DRC-Coder can be extended to a wide range of DRC-related applications.
For example, we can use our image analysis functions and include human interactive feedback in each Planner's response to realize a DRC-explanation chatbot.
We also aim to extend our framework to other areas of physical design that require multi-modal reasoning.
Finally, as DRC-Coder unlocks LLM's capability for a complex engineering task in EDA, we hope to stimulate future research on developing LLM-agents in this field.

\section*{Acknowledgement}
This work is supported in part by NVIDIA Corporation and NSF under Grant No. 2106828.